\documentstyle{elsart}

\begin{document}
\begin{frontmatter}

\title{A Topological Formulation of the Standard Model}

\author[cfa]{Marco Spaans\thanksref{em1}}

\address[cfa]{Harvard-Smithsonian Center for Astrophysics, 60 Garden
Street MS 51, Cambridge, MA 02138, USA}

\thanks[em1]{mspaans@cfa.harvard.edu}

\begin{abstract}
A topological theory for the interactions in Nature is presented.
The theory derives from the cyclic properties of the topological
manifold $Q=2T^3\oplus 3S^1\times S^2$ which has 23 intrinsic degrees of
freedom, discrete $Z_3$ and $Z_2\times Z_3$ internal groups, an SU(5) gauge
group, and an anomalous U(1) symmetry. These properties reproduce the standard
model with a stable proton, a natural place for CP violation and
doublet-triplet splitting. The equation of motion for the unified theory is
derived and leads to a Higgs field.
The thermodynamic properties of $Q$ are discussed and yield a consistent
amplitude for the cosmic microwave background fluctuations. The manifold $Q$
possesses internal energy scales which are independent of the field theory
defined on it, but which constrain the predicted mass hierarchy of such
theories. In particular the electron and its neutrino are identified as
ground states and their masses are predicted.
The correct masses of quarks
and the CKM mixing angles can be derived as well from these energy scales
if one uses the anomalous U(1) symmetry.

Furthermore, it is shown that if the Planck scale topology of the universe
involves loops as fundamental objects, its spatial dimension is equal to
three. The existence of the prime manifold $T^3=S^1\times S^1\times S^1$ is
then required for a dynamical universe, i.e.~a universe which
supports forces. Some links with M-theory are pointed out.\hfill\break\noindent
Keywords: general relativity --- quantum cosmology\hfill\break

\end{abstract}
\end{frontmatter}

\section{Introduction}

One of the outstanding questions in quantum cosmology and particle physics
is the unification of gravity with the electro-weak and strong interactions.
Much effort has been devoted in the past years to formulate a purely
geometrical and topological theory for both types of interactions[1,2].
Probably best known are the theories involving super-gravity and
superstrings[1]. These theories have been shown recently to be unified in
M-theory, although the precise formulation of the latter is not known yet.
Two outstanding problems in these approaches are compactification down to
four dimensions and the existence of a unique vacuum state.
This work aims at extending the topological dynamics approach presented in [3]
(paperI from here on) to include non-gravitational interactions and to
construct a Quantum-gravitational Grand Unified Theory (QGUT).

The main results of paperI are as follows:\hfill\break
The properties of space-time topology are governed by homotopically
inequivalent loops in the {\it prime} manifolds $S^3$,
$S^1\times S^2$, and $T^3=S^1\times S^1\times S^1$.
This is the only set which assures Lorentz
invariance and the superposition principle as formulated by the Feynman path
integral for all times. The dynamics of the theory are
determined by the loop creation ($T^\dagger$) and loop annihilation ($T$)
operators. On the Planck scale, space-time has the unique structure of a
lattice of Planck size three-tori with 4 homotopically
inequivalent paths joining where the $T^3$ are connected through
three-ball surgery. The existence of this four-fold symmetry
leads to SO(n) and SU(n) gauge groups as well as
the numerical factor $1/4$ in the expression for black hole entropy.
The number of degrees of freedom of the prime manifolds,
referred to as the prime quanta, under the action of
$O=TT^\dagger +T^\dagger T$, are 1, 3, and 7 for the three-sphere,
the handle manifold, and the three-torus, respectively.
During the Planck epoch, $S^1\times S^2$ handles (mini black holes) can
attach themselves to the $T^3$ lattice which can lead to interactions
between the degrees of freedom of the prime manifolds.
The $S^1\times S^2$ prime manifolds are referred to as charges on the $T^3$
lattice from hereon. Finally, it was found that the
cosmological constant $\Lambda$, as the spontaneous creation of mini black
holes from the vacuum, is very small and proportional to the number of
(macroscopic) black holes at the current epoch.
This result follows from the generalization of Mach's principle in which
local topology change is a consequence of global changes in the matter degrees
of freedom and vice versa.

This paper is organized as follows. Section 2 presents a derivation of the
dimension of the universe and its link with Lorentz invariance as well as the
necessity of a $T^3$ Planck scale topology for the existense of forces.
Section 3 discusses the construction of a QGUT based on the charged $T^3$
lattice and presents the derivation of the equation of motion.
Section 4 discusses the properties of
a quantized state space based on the equation of motion and the cyclic
properties of $Q$, which reproduces the standard model.
Section 5 contains the conclusions.

\section{The Dimension of the Universe, Lorentz Invariance and the Superposition Principle}

\subsection{Linked Loops and Spatial Dimensionality}

In paperI it is shown that a lattice of three-tori $L(T^3)$ supports
many homotopically distinct but otherwise equivalent paths between any two
points. This allows for a direct topological implementation of the superposition principle through the Feynman paths integral.
In this approach, the spatial dimension of the universe is postulated,
and ideally one would want to derive it from first principles.

Clearly, the invariance of the speed of light should hold in any quantum field
theory. Since the three-torus bounds a Lorentz four-manifold with SL(2;C) spin
structure (paperI), i.e.~is {\it nuclear} (like the handle),
the superposition principle on $L(T^3)$
implies Lorentz invariance in (3+1)-dimensional space-time. Conversely, the
loops in the prime manifolds are considered to be fundamental objects and as
such the homotopic structure of $T^3$ should be unaffected by dynamics.
Therefore, in a loop-topological sense, field interactions should correspond to
loops which are linked on the junctions of connected three-tori.
Such links preserve the underlying loop homotopy of the $T^3$ lattice.
Note that this implementation of the dynamics is quite different from string
theory, where the strings can break and merge.

It follows immediately that
one can exclude spatial dimensions $\le 2$ in the link picture.
Furthermore, any two linked loops embedded in a space of dimension $\ge 4$ are
homeomorphic to two separate loops, i.e.~no intersection/interaction need ever
occur in the deformation. Because Lorentz invariance implies time
dilation of the arbitrary intersection point of two loops, one wants the
linkage of two loops to be a topological invariant and to be independent of the
particular embedding. That is, the number of interactions should be both a
Lorentz scalar as well as an intrinsic property of the underlying topological
manifold which describes the dynamics. It follows that
the spatial dimension of a relativistic universe must be $D=3$.
As a corollary one finds that the occurrence of spin $1/2$ in the universe,
a property of the three-torus through the SL(2;C) spin structure it induces on
the four-manifold it bounds, follows from the superposition principle.

Conversely, the sum of the prime quanta, $d=1+3+7=11$, implies that the
kinematic dimension of the universe is eleven.
This dimension reflects the number of degrees of freedom in the fundamental
constituents (``basis vectors'') of space-time, the prime manifolds. That is,
$d$ reflects the
different ways in which the loop can exist in space-time on the Planck scale
when direct interactions between three-tori and handle are neglected (paperI).
This is elaborated in Section 3.

\subsection{The Three-Torus and the Existence of Forces}

The existence of the prime manifold $T^3$ should
ideally be derived from some fundamental physical principle. To find this
physical principle, the nature of gravitation will be investigated.
The equivalence principle in general relativity
leads to the notion of gravitational accelerations as
an expression of space-time curvature. In analogy with Einstein's thought
experiment one can ask how observers can determine the existence of a
{\it net} force
which acts on some object. Obviously, one wants to determine whether the object
is accelerating. To do this, one formally needs no less than three distinct
spatial measurements of the object's position. To extend such an experiment to
arbitrary small scales, the quantum-mechanical notion of position as an
expectation value comes into play. The view is taken here that even when
a particle is replaced by a field and a probability amplitude, the
concept of a force is equivalent to an interaction between (quantum) fields
and the application of Yang-Mills theory then leads to the notion of a field
strength or a curvature two-form generated by some gauge potential.

In general relativity, the existence of a force is interpreted as a curvature
of space-time. Mathematically, a curvature involves the second derivatives of
some metric field. Physically, the required existence of such an object is
a powerful constraint. The occurrence of a second derivative of any field
requires that field to be defined in three points. These three points need not
be infinitesimally close because the coarseness of a derivative only depends
on the measurement
scale one is interested in. This leads to a natural partition of space
into triplets. That is, whether there is a force field or not, its possible
existence requires any triplet of points to be viewed as a physical entity.

This yields a formal object $s$ denoted as
$s=\{\cdot\cdot\cdot\}\equiv [123]$, where the three dots indicate
the notion of a second derivative defined by three {\it arbitrary} points.
For interactions to occur and for space-time curvature to be induced by some
metric field, the object should be endowed with a topological structure
$\sigma$, yielding the dynamical object $S=\sigma (s)$. The structure $\sigma$
follows from the realization that all three points, as well as the paths
(partial trajectories) connecting them, are distinct. If the latter were not
the case, then on the Planck scale one could only superpose non-interacting
trajectories. This then implies
$$[1]\ne [2]\ne [3],\eqno(1a)$$
$$[12]\ne [23]\ne [13],\eqno(1b)$$
and
$$[ab]=[ba],\eqno(2)$$
where the last relation reflects that there is no preferred orientation on $S$.
The paths in relation (1b) should be topologically distinct to satisfy (1a).
Clearly, these relations imply the homotopic structure of the three-torus with
the loops $[12]$, $[23]$ and $[13]$.
Conversely, they do not specify the shape of $T^3$
or favor any loop in the three homotopic equivalence classes. Since the
force argument is independent of scale, the three-tori should be
inter-connected which immediately leads to the $T^3$ lattice of
paperI and the results of section 2.1. Finally, the above argument
does not provide any specific numerical value for the
Planck scale $\ell_{\rm Planck}$. If anything, all finite values, with equal
probabilities, are allowed except zero.

\section{Quantum Gravity, Superstrings and Topological Dynamics}

\subsection{QGUT Phenomenological Preliminaries}

In paperI it was shown that interactions between $T^3$ and $S^1\times S^2$
occur as matter traverses the wormholes which are attached to $L(T^3)$.
The three-torus has 3 homotopic
branches and it is easy to see that individual branches yield spin 1 particles
and pairs of branches lead to spin 1/2[4]. The handle introduces an additional
degree of freedom which can increase or decrease the homotopic complexity of
the structure to which it is attached, e.g.~provide single loop paths through
two connected loops or
bind two separate loops. Therefore, as a handle connects itself to two
singlets or connects two branches of a doublet, where singlets and doublets
should be viewed as different physical substructures of $T^3$, it can
effectively convert 1-bosons into 1/2-fermions and vice versa.
The aim is now to find a manifold consisting of three-tori and handles, which
provides the right number and properties of elementary fermions and bosons.

In paperI it is shown that the lattice junctions support SO(n) and
SU(n) symmetry groups because $L(T^3)$ naturally yields quadratic and quartic
interaction terms. Above it is found that the kinematic dimension of
the homotopic theory is $d=7+3+1=11$. The 1 in this formula corresponds to the
homotopically trivial manifold $S^3$ which is needed in the construction of
$L(T^3)$ and which provides the large scale topology of the universe. The
junctions which connect the three-tori correspond to the propagators in a
field theory and are homotopically equivalent to line elements. It follows
that the dimension of the junction groups which decribe the interactions
between the propagators on $L(T^3)$ is 10, i.e.~SU(5) and SO(10) which are
viable candidates for a GUT.

\subsection{QGUT Construction}

\subsubsection{The Fundamental Topological Manifold $Q$}

A manifold $Q=aT^3\oplus bS^1\times S^2$ which is built from three-tori and
handles should have an odd number of
constituents because only odd sums of nuclear primes bound Lorentz manifolds
(paperI). Because the supersymmetry is
implemented homotopically, the number of loops in the $a$ three-tori should be
equal to the number of mouths of the $b$ handles, i.e.~$3a=2b$.
Finally, any solution $R$ which has $a_1=na$ and $b_1=nb$ can be considered a
multiple of the smallest solution $aT^3\oplus bS^1\times S^2$. Since one wants
to construct a lattice $L(Q)$, the minimal solution is the desired one.
Therefore, the coefficients $a=2$ and $b=3$ result. This requires
the direct sum of three handle manifolds and two three-tori. In $Q$,
each handle is connected to two loops and a
homotopically supersymmetric manifold can be written as
$$Q=2T^3\oplus 3S^1\times S^2,\eqno(3)$$
which assures nuclearity and therefore Lorentz invariance. When the density of
mini black holes is large enough to support $Q$, the universe is said to be
$Q$-supersymmetric. As an aside it is
worthwhile to mention that the expression for $Q$ bears close resemblance to
the effective geometric string action (the Chern-Simons form)
$S_{\rm g}=2/3A\times A\times A + A\times DA$, for the gauge string field $A$
(viewed as a homotopic loop $S^1$) and the covariant derivative $D$. In fact,
the existence of the $T^3$ internal space seems to hint at a link with
heterotic string duality[1].

\subsubsection{The Equation of Motion for Constant Charge}

In paperI an elaborate derivation was given for the topological dynamics of
space-time in terms of the $T^3$, $S^1\times S^2$, and $S^3$ prime manifolds
when the loop is the fundamental object.
The derivation in paperI was based on 1) the existence of homotopic classes of
paths in a Lorentz invariant topological manifold as representatives of the
linear superposition principle, and 2) the algebra of loop creation and
annihilation operators which reflects the fact that the homotopy of space-time
is a dynamical quantity and is non-commutative in nature.
In order to develop a field theory which involves matter interactions at the
Planck energy, the topological object $Q$ is considered to be fundamental.
The equation of motion should then follow from some continuum limit of the loop
algebra acting on the four-manifold $Q\times R$ and the quantum fields defined
on it.

The homotopic equation of motion from paperI for a set of prime manifolds
${\cal T}_j$ in the limit of a continuous time variable is given by
$$\sum_j [T{\cal T}_i,T^{\dagger}{\cal T}_j]/{\cal T}_j=
(TT^{\dagger}+T^{\dagger}T){\cal T}_i.\eqno(4)$$
Because the topology is fixed to be that of $Q$ now, the interaction between
the prime manifolds on the left hand side in Equation (4), must be replaced by
the {\it self-interaction} of a field which lives on $Q\times R$.
Furthermore, the natural limit of the loop creation and annihilation operators
is that of two differential operators $\partial$ and $\partial^\dagger$
with space-time dimension four. $\partial$ and $\partial^\dagger$ must commute
in the continuum limit because $[T,T^\dagger]=1$ and the 1 on the right hand
side reflects only the discrete nature of the loop algebra. These differential
operators should also be conjugate in order to yield a scalar operator of the
form $\Box =\partial\partial^\dagger =\partial^\dagger\partial =
\partial_\mu\partial^\mu$, $\mu =1..4$.
Finally, the right hand side of Equation (4) reflects the fact that the prime
manifolds are fundamental objects and that their prime quanta cannot be
altered by the topological dynamics of space-time. Now that $Q$ is the
fundamental object and the dynamics of the field it supports are described by
the equation of motion, this right hand side should vanish.

This requires the identifications
$(T,T^{\dagger}\rightarrow\partial_\mu ,\partial^\mu)$,
${\cal T}_i\rightarrow q_\lambda$ for the left hand side of Equation (4).
This yields for the four vector $q_\lambda$ (see below)
$$q^\mu [q_\lambda\Box q_\mu -q_\mu\Box q_\lambda ]=0,\eqno(5)$$
One finds that the resulting {\it self-interaction is cubic}, as
demanded by the triple loop structure on the
three-torus, and linear in the second derivative operators.
The possible functional forms of $q_\lambda$ are determined by
the quadruplet solutions to the equation of motion, and the cyclic
structure of $Q$ which imposes boundary conditions.

The four-tensor $q$ is of rank one and its four components
correspond to the four currents, i.e.~wave amplitudes, which
can flow along the homotopically inequivalent paths associated
with the $T^3$ lattice (paperI). The square of the absolute value of this
object gives the probability to find the energy of $Q$, the dimensional number
$m_{\rm Planck}$ in the theory, concentrated in some point along any of the
four paths through $Q$. The individual components of $q_\lambda$ yield this
probability for each path. To normalize the probability distribution on the
underlying lattice of three-tori, one
should integrate over the discrete volume $V_{\rm Planck}$.

In PaperI it was shown that the SL(2;C) gauge group follows
naturally from a nuclear manifold. The theory is therefore manifestly
Lorentz invariant with $\partial^\mu =\eta^{\mu\nu}\partial_\nu$ and $\Box$
the d'Alambertian.
Obviously, the wave function $q_\mu$ is complex-valued and $q^\mu$ is defined
as $\delta^{\mu\nu}q^*_\nu$ to assure a real-valued inner product and therefore
positive definite probabilities.
A large class of solutions of (5) is determined by the wave
equations $\Box q_\lambda =0$. This solution is of the form
``the boundary of the boundary is zero'', in
analogy with the sourceless Maxwell equations. Its solutions are represented
by the well-known traveling wave forms. Another class of solutions is
determined by the Klein-Gordon equations $\Box q_\lambda +m^2q_\lambda =0$.
Because these solutions are applicable to a constant charge system, they
also hold for the neutral $T^3$ lattice.

\subsubsection{Low Energy Behavior}

The properties of the resulting QGUT depend on both the charge on
the $T^3$ lattice as well as the energy density of the
matter degrees of freedom. The former sets the topology of the Planck scale
manifold, whereas the latter determines which subgroups of the SU(5) symmetry
that lives on the junctions of $L(T^3)$ are realized.

Recently, it has been shown that M-theory unites many of the properties of
strings. Although no complete formulation of M-theory exists, it
exhibits many encouraging properties. In particular, the manifold
$T^3$ plays a fundamental role in the possible explanation of string duality.
It is therefore tempting to suggest that for vanishing charge and large
energy (above the GUT scale), M-theory results on $L(T^3)$.
That is, M-theory corresponds to the neutral quadruplet solutions
on the $T^3$ lattice. Clearly, these results hint at underlying properties
which remain to be found. The aim of this work will therefore be to use the
homotopic properties of $L(T^3)$ and $Q$ to find a formulation of the standard
model with particular emphasis on the number and masses of elementary
particles.

\subsubsection{The Equation of Motion in the Presence of Charge Fluctuations}

For large charge densities, $Q_{\rm h}\approx 1$, $L(T^3)$ is completely
interconnected by handles. In this limit $L(T^3)\rightarrow L(Q)$ as the Planck
scale manifold for $t\sim t_{\rm Planck}$. Unlike the three-tori, the mini
black holes couple directly to the matter degrees of freedom through the
processes of accretion and Hawking radiation. Therefore, even though the charge
density is of the order of unity, handles are continuously being created as
described in paperI and destroyed through evaporation and merging.

These quantum perturbations in the local number and physical properties of
handles lead to the generation of an additional field. This field is envisaged
to reflect phase changes in the currents, i.e.~the wave amplitudes
$q_\lambda$, flowing through $L(T^3)$. The fundamental object to solve for on
$L(Q)$ is therefore $\Omega_\lambda\equiv {\rm e}^{2\pi i\phi}q_\lambda$, with
$\phi$ a function of time and position. The real scalar field $\phi$
introduces a modulation of the wave amplitude as a driving force would for a
harmonic oscillator. This phase transformation leads to the full QGUT equation
of motion
$$4\pi i\partial_\nu\phi [(\partial^\nu q_\lambda )q^\mu q_\mu -(\partial^\nu
q_\mu )q^\mu q_\lambda ]=q_\lambda q^\mu\Box q_\mu -q^\mu q_\mu \Box
q_\lambda ,\eqno(6)$$
with the scalar constraint
$$q^\mu q_\mu ={\rm cst}.\eqno(7)$$
Due to the continuous creation and destruction of $\Theta$ manifolds, it is
possible to travel from a path to any other path. Therefore, from the point of
view of the wave amplitudes, the points along a path become indistinguishable,
although the homotopy persists.
The scalar constraint (7) then expresses the fact that the {\it total}
probability to find
all the mass-energy of $Q$ in some point is the same for all points in $Q$,
albeit at the expense of the scalar field $\phi$. This also implies that both
on $L(Q)$ and $L(T^3)$ the solutions are identical on every three-torus.

Because the evolution of $\phi$ is driven by the handle degrees of freedom,
it follows that on the neutral $T^3$ lattice the field $\phi$ obeys the
limiting condition $\partial_\nu\phi =0$, and is effectively frozen in.
In the low
charge limit, the field $\phi$ therefore has a constant, Lorentz invariant
vacuum expectation value $<0|\phi (x)|0>=c\not =0$, defined on the junctions
of the $T^3$ lattice. A non-zero vacuum expectation value of $\phi$ requires
$\mu^2<0$ and leads to spontaneous symmetry breaking, as first suggested by
Nambu and co-workers. The additional Poincar\'e scalar $\phi$ can
therefore be identified with a Higgs field, although the number of Higgses
is not constrained. In fact, upon freeze-out only the total norm $\phi_l\phi^l$
would be fixed. Furthermore, the constraint (7) no longer applies since
$\Theta$ has evaporated.

\subsubsection{The Heat Capacity of $Q$}

The number of degrees of freedom of $Q$ under the action of the loop algebra
is defined as
$$N_Q=(TT^\dagger +T^\dagger T)(2T^3\oplus 3S^1\times S^2)=23.\eqno(8)$$
In the loop homotopic approach adopted here, these degrees of freedom are
all equivalent and they reflect the different ways in which the loop creation
and annihilation operators can act on $Q$ while conserving the homotopic
structure of the topological manifold. Any fields which are defined on the
non-prime manifold $Q$ are therefore automatically partitioned over these 23
degrees of freedom.

For the neutral submanifold $P=T^3\oplus T^3$,
one has $N_P=14$. The latent heat associated with the evaporation of the
handle triplet $\Theta =3S^1\times S^2$ is therefore
$$H=(N_Q-N_P)m_{\rm Planck}/N_Q=9m_{\rm Planck}/23.\eqno(9)$$
This number is uniquely determined by the homotopic structure of space-time
and the Planck mass. Since both three-tori in the structure $Q$ are equivalent,
the specific heat per three-torus $h$ is given by
$$h=H/2=9m_{\rm Planck}/46.\eqno(10)$$

\subsubsection{The QGUT Phase and Large Scale Structure}

From the above discussion it follows that the QGUT energy scale corresponds to
a size and matter density of the universe where mini black holes are
formed rapidly enough to sustain the topological manifold $Q$.
During this phase the 23 degrees of freedom of $Q$ are accessible to the
wave amplitudes $q_\lambda$ traveling through $L(Q)$. As these currents
self-interact, they do so cubically on the three-torus as in Equation (6).
The currents carry the mass-energy of the universe and their interactions
determine the perturbations in the mass-energy associated with the 23
equivalent degrees of freedom.

The strength of these perturbations in the mass-energy is given by
$$\delta\rho /\rho(Q) =N_Q^{-3}=8.2\times 10^{-5}.\eqno(11a)$$
This amplitude is the first non-zero correction term to the average energy
in any of the degrees of freedom of $Q$ on purely thermodynamic grounds.
PaperI discusses the effective dispersion of a large ensemble of Gaussian
perturbations on a lattice of three-tori and this leads to the $1\sigma$
effective amplitude in the mass-energy $E$
$$\delta\rho /\rho(E) =3.7\times 10^{-5}.\eqno(11b)$$
For adiabatic perturbations on the horizon scale the fluctuations in the Cosmic
Microwave Background temperature is $1/3$ of $\delta\rho /\rho(E)$ and one
finds
$$\delta T/T={{1}\over{3}}\delta\rho /\rho(E) =1.2\times 10^{-5}.\eqno(12)$$
This amplitude is consistent with recent COBE measurements.

\section{The Standard Model from the Symmetries of $Q$}

With the fundamental manifold and the equation of motion which lives on it
in place, one now needs to isolate the symmetry properties of $Q$ and solutions
to Equations (5) and (6) in order to arrive at a description of the standard
model with as few free parameters as possible.

\subsection{Preliminaries}

\subsubsection{Discrete Groups Generated by $P$ and $\Theta$}

The handle manifolds which are
created in quantum fluctuations always form triplets on $L(T^3)$.
Therefore, the effective action $s^3$, of the
triplet as a whole, on any excitation of $Q$ obeys
$$s^3=1.\eqno(13)$$
That is, a round trip along the manifold $\Theta$ necessarily picks up three
phases, which should add up to $2\pi$ since the loop algebra satisfies
$[T,T^\dagger ]=1$. Because all 3
handles are identical, this implies a $Z_3$ invariance for the individual
quantum fields in the theory defined on $Q$ with angles
$\theta_i=\{0,\pm 2\pi /3\}$. From the same arguments it follows that the
submanifold $P=T^3\oplus T^3$ in $Q$ generates a $Z_2\times Z_3$ symmetry
because one can distinguish neither the three loops in a three-torus nor a
$T^3$ contained in $P$. Also, the $Z_3$ groups belongs to SU(3) because the
latter consists of elements like (13).

\subsubsection{$T^3$ Junctions and U(1) Factors}

The important distinction between $T^3$ and $L(T^3)$, or $L(Q)$ for that
matter, is the presence of junctions which connect the individual three-tori
through
three-ball surgery and create a lattice. In paperI it was suggested that the
presence of a lattice facilitates a geometric description of gravitational
effects because the three-dimensional junctions can bend according to some
curvature tensor. Indeed, on scales much larger than $\ell_{\rm Planck}$ the
neutral lattice $L(T^3)$ appears as a smooth manifold. On the Planck scale on
the other hand, the existence of junctions between the three-tori generates
an additional U(1) symmetry. As one patches the individual three-tori together,
there is an arbitrary rotation, or twist, one can
perform without changing the topological properties of the manifold. Obviously,
there is only one twist per $Q$ manifold, i.e.~per junction, but each
individual three-torus in the lattice formally has six of them.
These additional U(1)
factors have been proposed as a possible resolution of the doublet-triplet
splitting problem[5]. From the discussion above it follows that the existence
of the $Z_2\times Z_3$ cyclic group through $P$ and the absence (presence) of
an anomalous U(1) sector are equivalent properties of the $Q$ ($T^3$) lattice.

\subsection{Construction of the Standard Model}

With these preliminaries in mind, the aim now is to reproduce the features
of the standard in the low energy
and low charge limit. The philosophy is to group the 23 homotopic degrees of
freedom of $Q$ into different equivalence classes under the action of the
scalar operators $A\equiv TT^\dagger$ and $B\equiv T^\dagger T$ on $P$ and
$\Theta$.

\subsubsection{Equivalence Classes and Broken $Q$-Supersymmetry}

The homotopic properties of $Q$, the SU(5) gauge group on the junction
and the $Z_3$ and $Z_2\times Z_3$ cyclic symmetries of $\Theta$ and $P$ will
lead to specific particle sectors. In this, the photon and graviton are not
viewed as being generated through the homotopic structure of $Q$, but result
from the junction degrees of freedom, i.e.~the U(1) twist and GL(4)
curvature of $L(T^3)$. Furthermore, there is a symmetry breaking
Higgs field which lives on the junction and can become massive, leading to
Higgs bosons.

The number of degrees of freedom $N_Q$ is the eigenvalue of the operator
$O=TT^\dagger +T^\dagger T\equiv A+B$ acting on $Q$.
There is a natural division of the 23 degrees of freedom under
$AQ=14Q$ and $BQ=9Q$. Furthermore, the decomposition $OQ=O(P\oplus\Theta )$
has the same distribution of degrees of freedom under $A$ and $B$ and leads to
the further divisions
$$AQ=OP=(8+6)P,\eqno(14)$$
with eight gluons and six quarks and
$$BQ=O\Theta =(3+6)\Theta ,\eqno(15)$$
with six leptons and three vector bosons. This identification follows from the
fact that the junction potential on $P$ supports the symmetry group SU(5).
The $P$ and $\Theta$
sectors decompose $Q$ and are therefore associated with subgroups of SU(5).
These subgroups can only contain SU($n\le 5$) and U(1) because of the junction
potential and twist. For SU(5)=SU(3)$\times$SU(2)$\times$U(1)
these constraints are satisfied. Since SU(3) contains 8 field particles
and SU(2) only 3, the identification of $P$ with QCD and $\Theta$ with the
electro-weak interaction is immediate.

A priori both fermionic and bosonic sectors exist for these equivalence
classes. That is, because the form of $Q$ is motivated by Lorentz invariance
and $Q$-supersymmetry, the identified equivalence classes can be both fermionic
and bosonic in nature.
When $Q$-supersymmetry is broken the structure $L(T^3)$
with the three-torus as fundamental Planck scale object assures Lorent
invariance. Subsequently, interactions are mediated by field particles which
travel along the 6 junctions surrounding any $T^3$. Furthermore, it is the
discrete three-torus with its seven degrees of freedom under the operator $O$
which supports a particle and its field quanta. Any field dynamics on $L(T^3)$
therefore requires the interaction of two field particles on a three-torus. If
these field particles are fermionic, this violates the Pauli exclusion
principle.

Thus, only bosonic field particles can carry the strong and electro-weak force,
and satisfy the Pauli exclusion principle on $L(T^3)$,
after the handle triplets
have evaporated. Therefore, even though the principle of supersymmetry is
necessary to identify the fundamental manifold $Q$, the requirement of Lorentz
invariance leads to $L(T^3)$ and forces field particles to be bosonic. As such,
it determines the way in which $Q$-supersymmetry is broken when the handle
triplet evaporates. The origin of the Pauli exclusion principle follows from
the homotopic structure of $T^3$ and the fact that a spin $1/2$
particle requires two loops on a three-torus for its support. For two identical
fermions this implies the general relation (see also \S 2.2 above)
$$[ac][cb]=[ca][ab],\eqno(16a)$$
which yields
$$[cb]=[ab].\eqno(16b)$$
This indicates that because one $S^1$ loop is a part of both fermions, the
other two are collapsed to one. The consequence is that the
three-torus becomes indistinguishable from the prime manifold
$S^1\times R_1$, with $R_1$ a Riemann surface of genus one. This manifold is
not nuclear and therefore breaks Lorentz invariance. It is straightforward to
verify that the definition of two identical bosons preserves the homotopy of
$T^3$ because they involve disjunct loops.

\subsubsection{Interpretation of the Equation of Motion}

Equations (5), (6) and (7)
describe the quantum-mechanical interactions of matter in full,
i.e.~including quantum gravity.
The boundary conditions for the solutions should
follow from the cyclic properties of $L(T^3)$.
The topology of $T^3$ requires the solutions $O_\lambda (x,y,z,t)$
to be periodic on a cube of size $L$ for every time $t$,
$$O_\lambda (x,y,z,t)=O_\lambda (x+L,y+L,z+L,t).\eqno(17)$$
The handle triplet then requires a solution which limits to $\phi =1/3$ when
$Q$-supersymmetry is broken. This phase relation
is consistent with the fact that the equation of
motion (6) is invariant under the global transformation
$\phi\rightarrow\phi +\alpha$. This freedom is
fixed by the underlying topology of $Q$.
The initial conditions at $t=0$ for the solutions of (6) can then be taken
as $q_\lambda (0)={\rm cst}$, derivatives $\partial_\nu q_\lambda (0)$
given by the Planck energy and $\phi (0)=0$.

For constant values of $\phi$, i.e.~$Q_{\rm h}\sim 0$, the general solution of
(6) can naturally limit to the Klein-Gordon equations
$(\Box +m^2)q_\lambda =0$, which support massive particles. This situation
applies when $Q$-supersymmetry is broken. In this phase,
the statistics of the mass-energy distribution is represented by the
absolute value $q^\mu q_\mu$ of the wave modes.
To follow the evolution of the matter degrees of freedom during this epoch,
one should solve Equation (5) with the QGUT end
solution of (6) as initial conditions.
Once, the GUT is broken at some energy (see below) the Einstein equation
describes the later time evolution of the mass-energy distribution.

A question which can be
assessed is the nature of the statistics of mass-energy
fluctuations after the GUT is broken. For this, the solution space
of Equation (5) needs to be investigated.
Finally, the cubic periodicity of the solutions to the equation of motion on
$L(T^3)$ does not lead to a global $T^3$ topology at the present epoch. The
Planck size branches of the three-torus are only accessible to sufficiently
energetic particles and the global appearance of the universe is therefore
dominated by the $S^3$ junctions (paperI).

\subsubsection{Symmetry Groups, the $\mu$ Problem and Proton Stability}

The homotopic theory as it stands identifies SU(5) and SO(10) as gauge groups
and $Z_3$ and $Z_2\times Z_3$ as additional discrete groups, but leaves room
for possible (and in fact necessary) extensions of the minimal standard model.
A fundamental problem in supersymmetric GUT is the doublet-triplet splitting
problem which results from the unavoidable mixing of Higgs doublets
$H,\bar{H}$ with their colored triplet partners $T,\bar{T}$.
This also leads to an unacceptably rapid proton decay. In [6] it was suggested
that there is no need for the heavy triplet if its Yukawa coupling constant is
strongly suppressed with respect to the one of the doublet. This mechanism
requires an SO(10) invariant operator with tensor indices $i,k$
$${{Y_{\alpha ,\beta}}\over{M_{\rm GUT}}}10_i45_{ik}16^\alpha
\gamma_k16^\beta ,\eqno(18)$$
in which $16^\alpha$ ($\alpha =1..3$) are three families of matter fermions,
$10_i$ ($i=1..10$) is the multiplet with $H,\bar{H}$ ($i=7..10$) and
$T,\bar{T}$ ($i=1..6$). The 45 is the GUT Higgs in the adjoint presentation of
SO(10), $Y_{\alpha ,\beta}$ is the coupling constant matrix, and the $\gamma_i$
denote the matrices of the SO(10) Clifford algebra. To realize this effective
operator, the 10-plet must transform under the symmetry group $Z_2\times Z_3$
such that the 10-plet does not couple to the GUT Higgses and the 10-plet is
allowed to interact with $16^\alpha$ only in combination with the 45-plet[6].

A possible resolution of the $\mu$ problem in the standard model thus naturally
involves the cyclic group $Z_2\times Z_3$ associated with $P$. In addition,
the submanifold $\Theta$ generates $Z_3$. If one introduces a light gauge
singlet superfield $N$, then the triple interactions on the individual
three-tori fix the associated superpotential to be of the form
$$W=\lambda_1 N10^2+\lambda_2N^3.\eqno(19)$$
This potential is invariant under $Z_2\times Z_3$ because the singlet $N$
has an $Z_2$
invariance on $P$. Both $N$ and 10 do not transform under $Z_3$ due to
$\Theta$, and therefore they decouple from the heavy GUT Higgs fields.
This provides a natural resolution of the doublet-triplet problem in terms of
the homotopy of space-time as already anticipated in [6], {\it if} one accepts
the existence of an additional gauge singlet. In fact, the existence of this
object is a direct consequence of the anomalous U(1) symmetry which is hidden
in $Q$ and emerges when $Q$-supersymmetry is broken to $L(T^3)$ (see \S 4.1.2
above). Because the triplets have no coupling at all, it follows that the
proton is essentially stable even if the decoupled triplet is as light as its
doublet partner. That is, its decay rate is suppressed by a factor which is
no larger than $(M_W/M_{\rm GUT})^2$, with $M_W$ the mass of the weak scale.
Specific models based on SO(10) are
constructed in [6], and the existence of long-lived $T,\bar{T}$
supermultiplets in the 100 GeV-TeV mass range is discussed there as well.

\subsubsection{The Fermion Mass Hierarchy in the Standard Model}

Another fundamental problem which requires a resolution in GUT is the specific
form of the fermion mass hierarchy. The popular approach is to use the
anomalous U(1) gauge symmetry as a horizontal symmetry[5]. The
motivation is that the anomalous U(1) symmetry breaking scale is given by
$${{\surd\xi}\over{M_{\rm Planck}}}\sim 0.1-0.01,\eqno(20)$$
with $M_{\rm Planck}=2\times 10^{18}$ Gev the reduced Planck mass and $\xi$
the Fayet-Iliopoulos term[5]. This ratio
is of the order of the fermion mass ratios in neighbouring families.
The precise value of $<N>=\surd\xi /q$, with $q$ the anomalous U(1) charge,
depends on the value of
trQ=trQ$_{\rm obs}$+trQ$_{\rm hid}$, with contributions from the observable
and hidden matter singlet[5].
A second mass scale corresponds to $M_{\rm GUT}=<\phi >$,
which marks the energy at which the SU(5) theory is broken down
to the SU(3)$\times$SU(2)$\times$U(1) symmetry of the standard model.
Obviously, both energy scales should have a common origin.

When the handles evaporate, the expectation value of $\phi$ becomes $1/3$
which breaks the SU(5) symmetry. Since $L(T^3)$ is the central
object for a grand unified theory, the GUT energy scale should correspond to
$1/3$ times the specific energy per three-torus of a degree of freedom in $Q$.
That is, the breaking of a GUT corresponds to the fact that the energy
associated with the fundamental loop, which generates a non-trivial homotopy,
is no longer available to excite the loop creation and annihilation operators.
At this point, the wave amplitudes $q_\lambda$ can no longer be influenced by
Equation (5).
Because the 23 degrees of freedom on $Q$ are equivalent one expects approximate
equipartition of energy for the possible wave amplitudes on $Q$. Therefore,
one finds $M_{\rm GUT}=1/3m_{\rm Planck}/2N_Q\approx 8.8\times 10^{16}$ GeV,
with $m_{\rm Planck}=1/\surd G=1.22\times 10^{19}$ GeV in units with
$\hbar =c=1$. Above it
was shown that the latent heat per $T^3$ associated with breaking of
$Q$-supersymmetry, equals $h=9/46m_{\rm Planck}$. Therefore, one finds for the
singlet superfield $<N>=1/3h\approx 8.0\times 10^{17}$ GeV.
Note that this expectation
value is again purely topological. It should be emphasized that the
presence of the U(1) factors and the various energy scales are unique
predictions of the model, but that a number of field theoretic implementations
are possible.

Because $N$ is an SU(5) singlet, one normally finds identical mass matrices for
the charged leptons and the down quarks. Observationally, there is about a
factor of 3 splitting between the down quark and lepton masses in the same
family. In the homotopic theory presented here, the electro-weak and QCD
sector are associated with $\Theta$ and $P$, respectively. Because the degrees
of freedom of each sector are in equipartition with one another for
$Q_{\rm h}\sim 1$, the mass separation $\rho=M(\Theta )/M(P)$ on
$L(T^3)$ is given by the eigenvalues of the topological operator $O$. One finds
$\rho =9/14\approx 0.64$ for the ratio of the degrees of freedom.

The unique ratio $\delta =<N>/m_{\rm Planck}\approx 6.5\times 10^{-2}$
then yields $\epsilon =\rho\delta \approx 4.2\times 10^{-2}$ which reproduces
the observed pattern of quark masses in a supersymmetric SU(5) model as
in [5] with the anomalous U(1) symmetry of $L(T^3)$, according to
$$m_t:m_c:m_u\sim 1:\epsilon^2:\epsilon^4\eqno(21a)$$
$$m_b:m_s:m_d\sim 1:\epsilon :\epsilon^2\eqno(21b)$$
with the CKM mixing angles
$$\sigma_{12},\sigma_{23}\sim\epsilon ,\qquad s_{13}\sim\epsilon^2.\eqno(22)$$
The observed value of trQ is consistent with this result for $\epsilon$ if
there is a significant negative contribution from the hidden sector[5].

\subsubsection{The Absolute Scale of the Mass Hierarchy}

The mass scalings have been identified above for a particular implementation
of the anomalous U(1) symmetry, but no absolute scale has
been determined. It will now be shown that, just like the energy scales of the
fields generated by $Q$ are uniquely determined by the homotopic theory, so is
the low mass end of the lepton mass hierarchy. That is, ground state energies
can be identified for the electron and neutrino in the electro-weak sector,
because $Q$-supersymmetry is broken by the evaporation of $\Theta$.

Before $Q$-supersymmetry is broken, the 23 degrees of freedom of $Q$ are in
equilibrium and the configuration space ${\cal C}$ of possible (equivalent)
states on $Q$ consists of
$$X_Q=23!\approx 2.6\times 10^{22}\eqno(23)$$
elements. These states are associated with particles created
after symmetry breaking
and their mass is thus $m_{\rm Planck}/X_Q$. Because
$Q$-supersymmetry is broken when the handles evaporate, the particle in
question belongs to the electro-weak $\Theta$ sector. It should be charged
because $Q$ contains a U(1) sector (twist), and be a ground state since $Q$
generates the maximal size configuration space. It must therefore
fix the low part of the charged lepton spectrum. It follows that
$$m_{\rm e}^0=m_{\rm Planck}/X_Q\approx 0.47\quad {\rm MeV}\eqno(24)$$
determines the electron mass. This estimate agrees with observations at the
8\% level.

For the neutral
submanifold $P$ one has $X_P=14!$, which fixes the number of neutral substates
on $L(T^3)$ after $Q$-supersymmetry breaking, and yields the
extended configuration space ${\bar{\cal C}}$ with
$X_QX_P\approx 2.3\times 10^{33}$
elements. The energy of the neutral (neutrino) ground state of $\bar{\cal C}$
thus follows from
$$m_{\nu_{\rm e}}^0=m_{\rm Planck}/(X_QX_P)\approx 5.4\times 10^{-6}\quad
{\rm eV}.\eqno(25)$$
Because the neutrinos have no charge, they cannot be distinguished on $P$,
unlike the charged leptons which couple non-trivially to the U(1) sector on
the junction of $P$.
The photon and graviton are massless because they do not depend on the homotopy
of space-time. The masses of the vector bosons are determined by the breaking
of the SU(2) symmetry which requires the direct intervention of a Higgs field.

\subsubsection{Higher Order Corrections to the Mass Hierarchy}

Higher order quantum corrections on $L(T^3)$ occur, which correspond to
the cubic interactions of the particle wave
modes in their respective homotopic equivalence classes.
If $s_Q$ describes the dispersion of the probability distribution
on $Q$, then one can ask with what accuracy ${\cal A}$ the properties of $P$
can be determined, given it has a dispersion $s_P$. The uncertainty relation
then yields $s_Q={\cal A}s_P$. This question is relevant to the uncertainty
of the energy levels of $\bar{\cal C}$ since the evaporating mini black holes
occupy only a part of the total number of degrees of freedom on $Q$.
The relative uncertainty in the ground
state energies is therefore ${\cal A}=(14/23)^3=0.23$. From a
thermodynamic point of view this correction reflects a shift in the zero
point energy. Still, the topological degrees of freedom set an absolute lower
limit to the energy level and the shift can only be upwards. That is, the
uncertainty implies a lack of information or smaller entropy and therefore
a higher excited state. Clearly, given an energy $m_{\rm Planck}$, the entropy
of the universe can be maximized
by allowing for as many occupied states as possible. The accuracy ${\cal A}$ is
consistent with the {\it underestimate} for the electron mass.
Even though the statistics of the specific realizations of universes is not
known, one can give the confident limit on the electron mass
$$0.47<m_{\rm e}<0.58 \quad {\rm MeV},\eqno(26)$$
which is in good agreement with the measured value of $0.51$ MeV.

\subsubsection{CP Violation}

There is one final global degree of freedom on $L(T^3)$: the lattice junction
in a pair of three-tori can vibrate.
The energy $F$ of the excitation is given by
$$F={{m_{\rm Planck}-H}\over{X_QX_P}}\approx 3.3\times 10^{-6}\quad {\rm eV}.\eqno(27)$$
The denominator reflects the total number of configurations in $\bar{\cal C}$.
Until the mini black holes evaporate and the latent heat $H$ is released, a
typical energy of $m_{\rm Planck}-H$ is confined to the internal degrees of
freedom of $L(T^3)$.
This introduces an additional energy state in $\bar{\cal C}$ which cannot be
massive because it is below all particle thresholds, and is therefore
associated here with the lattice itself.

To excite this internal degree of freedom one requires a system of two
neutral particles which are in resonance. That is, (1) the particles should
not be independent because of a common decay route. This allows them to form a
mixture with two superposition states. (2) The difference in self-energy
between these superposition states should
be larger than $F$ to excite the vibration but smaller than the lowest particle
ground state in $\bar{\cal C}$.
It is well known that the difference in weak self-energy determined by
the superposition states
$${\rm K}_S\leftrightarrow 2\pi\leftrightarrow {\rm K}_S\eqno(28)$$
and 
$${\rm K}_L\leftrightarrow 3\pi\leftrightarrow {\rm K}_L\eqno(29)$$
for the decay of the neutral K$^0$ and $\bar{\rm K}^0$ mesons,
is extremely small and equal to $f\approx 3.5\times 10^{-6}$ eV.
Indeed, $f>F$ but below particle threshold.
It is this small asymmetry which allows the CP (but not CPT) violating K meson
decay processes to occur through
${\rm K}_L\rightarrow {\rm K}_S\rightarrow 2\pi$.

\section{Conclusion}

It has been shown that the requirement of Lorentz invariance follows from the
superposition principle and leads to a spatial dimension of the universe equal
to three. The presence of forces has been found
to imply the physical necessity of the three-torus as the fundamental Planck
scale object. Together with mini black holes these three-tori can form a
fundamental manifold which is Lorentz invariant and provides a natural
mechanism for symmetry breaking through black hole evaporation.
An equation of motion has been derived for a QGUT on this
supersymmetric manifold $Q=2T^3\oplus 3S^1\times S^2$, which naturally leads
to a Higgs field. The manifold $Q$
contains the necessary symmetry groups
for the standard model with a stable proton, CP violation and doublet-triplet
splitting, and possesses intrinsic energy scales which reproduce the cosmic
microwave background fluctuations.
These properties are independent of possible extensions of the standard model,
which will allow tighter constraints to be placed on GUTs.

The author is indebted to G.~van Naeltwijck van Diosne, J.A.A.~Berendse-Vogels,
W.G.~Berendse and M.A.R.~Bremer for valuable assistance.

\end{document}